\documentclass[usenatbib]{mn2e}
\usepackage{bm,amssymb}
\usepackage{graphicx}
\usepackage{aas_macros}

\newcommand{\kms}{{\rm \, km \, s^{-1}}}
\newcommand{\msun}{\rm{M}_{\odot}}

\title[AGN-Driven Galactic Winds]
{Galaxy-Scale Outflows Driven by Active Galactic Nuclei}
\author[J. DeBuhr et. al.]{Jackson~DeBuhr,$^1$ Eliot~Quataert,$^{1,2}$ and Chung-Pei~Ma$^2$ \\
  $^1$Department of Physics, University of California, Berkeley, CA 94720, USA \\
  $^2$Department of Astronomy and Theoretical Astrophysics Center,
  University of California, Berkeley, CA 94720, USA}

\begin{document}

\maketitle

\begin{abstract}

We present hydrodynamical simulations of major mergers of galaxies and study the effects of winds produced by active galactic nuclei (AGN) on interstellar gas in the AGN's host galaxy.  
We consider winds with initial velocities $\sim 10,000 \kms$ and an initial momentum (energy) flux  of $\sim \tau_w L /c$  ($\sim 0.01 \, \tau_w \, L$), with $\tau_w \sim 1-10$.   The AGN wind sweeps up and shock heats  the surrounding interstellar gas, leading to a galaxy-scale outflow with velocities $\sim 1000 \kms$, peak mass outflow rates comparable to the star formation rate, and a total ejected gas mass $\sim 3 \times 10^9 \, M_\odot$.  Large momentum fluxes, $\tau_w \gtrsim 3$, are required for the AGN-driven galactic outflow to suppress star formation and  accretion in the black hole's host galaxy.  Less powerful AGN winds ($\tau_w \lesssim 3$) still produce a modest galaxy-scale outflow, but the outflow has little global effect on the ambient interstellar gas. 
We argue that this mechanism of AGN feedback can plausibly produce the high velocity outflows observed in post-starburst galaxies and the massive molecular and atomic outflows observed in local ultra-luminous infrared galaxies.   Moreover, the outflows from local ultra-luminous infrared galaxies are inferred to have  $\tau_w \sim 10$, comparable to what we find is required for AGN winds to regulate the growth of black holes and set the $M_{BH}-\sigma$ relation.   We conclude by discussing theoretical mechanisms that can lead to AGN wind mass-loading and momentum/energy fluxes large enough to have a significant impact on galaxy formation.
   
\end{abstract}

\section{Introduction}
\label{sec:intro}

Accretion onto a central massive black hole (BH) in a galactic nucleus produces energy in the form of radiation, relativistic jets, and wider angle (less-collimated) non-relativistic ($v \sim 10^4 \kms$) outflows \citep{krolik99}.   The coupling of this energy output to gas in galaxies and in the intergalactic medium is believed to play an important role in galaxy formation, potentially regulating the growth of massive galaxies and the thermal properties of the intracluster medium in galaxy groups and clusters (e.g., \citealt{silk98,croton06}).

The impact of this `feedback' on the gas {\em in} galaxies is particularly uncertain, both because the interstellar (ISM) gas is denser, and thus more difficult to affect dynamically, and because much of the ISM subtends a relatively modest solid angle relative to a central active galactic nucleus (AGN).   However, analytic estimates and numerical simulations have demonstrated that if a modest fraction of the energy produced by accretion onto a central BH can couple to the surrounding gas, it can unbind the interstellar gas (e.g., \citealt{silk98,dimatteo05}).  The physical processes most likely to produce such an effect are winds \citep{king03, king11}, radiation pressure \citep{murray05}, and/or Compton heating \citep{sazonov04} from a central AGN.   Understanding how this works in detail is one of the major challenges in our understanding of the connection between AGN physics and galaxy formation.

In this paper, we assess the influence of AGN winds on gas in the AGN's host galaxy using three-dimensional numerical  simulations.   Previous analytic work and one and two-dimensional simulations have demonstrated that  AGN winds can in principle sweep up and drive gas out of galaxies, potentially explaining the $M_{BH}-\sigma$ relation and the dearth of gas and ongoing star formation in massive, early-type, galaxies (e.g., \citealt{king05,king11,novak10,ostriker10} and references therein).    
 
Observationally, there is strong evidence that AGN indeed drive powerful outflows.    Broad-absorption line (BAL) quasars, which show blue-shifted absorption lines in the rest-frame ultraviolet with inferred outflow velocities $\sim 10,000-40,000 \kms$, represent over $\sim 40\%$ of quasars in infrared selected samples \citep{dai08}. A similar fraction of radio-quiet quasars show evidence for high velocity outflows in X-ray absorption line spectroscopy  \citep{tombesi10}.  It is likely that {\em all} quasars possess such outflows but that they are only observed when the system is viewed modestly edge-on \citep{murray1995}.   However, determining the mass-loss rate from spatially unresolved absorption-line observations is notoriously difficult given uncertainties in the radius of the absorbing gas.
In a handful of low-ionization BAL quasars (in particular, FeLoBALs) this degeneracy has been broken, suggesting mass loss rates significantly larger than the black hole accretion rate \citep{moe09, bautista10,dunn10,claude11}.   These observations trace absorbers at large distances from the BH ($\sim$ kpc), in contrast to most of the high ionization UV and X-ray absorption seen in BAL quasars, which arises at $\lesssim 1$ pc.   In addition to these well-characterized outflows, it is possible that quasars drive even more powerful outflows during phases of super-Eddington accretion \citep{kingpounds03, king11} or during obscured phases when the AGN's radiation  is trapped in the galactic nucleus, enhancing the force on the ambient gas \citep{debuhr10, debuhr11}.

To characterize the impact of AGN winds on their host galaxy, we carry out numerical simulations of major galaxy mergers with models for BH growth and feedback.   Although it is by no means certain that all quasars are associated with mergers, this provides a convenient and well-posed model in which to study gas inflow in galactic nuclei.  It also allows us to readily compare our results to the extensive previous literature on AGN feedback during mergers.   

In the next section (\S \ref{sec:methods}) we summarize our methods, emphasizing our implementation of AGN winds and our treatment of ISM cooling.   We then describe our key results in \S \ref{sec:results}. We conclude in \S \ref{sec:discussion} by discussing the implications of our results for models of AGN feedback and for galactic winds driven by AGN.   We also compare our results to observations of high-velocity outflows from post-starburst galaxies and ultra-luminous infrared galaxies and summarize the theoretical processes that can produce the large mass-loading and momentum fluxes we find are necessary for AGN winds to have a substantial impact on the surrounding ISM.   

\section{Methodology}
\label{sec:methods}

We use a modified version of the TreeSPH code GADGET-2 \citep{springel05} to
perform simulations of equal-mass galaxy mergers.  This version of the code 
includes the effective star formation model of~\cite{springel03} 
(hereafter SH03) and the radiation pressure AGN feedback model of 
\cite{debuhr11} (hereafter DQM).  As described below, we modified the code further to implement a model of feedback via AGN winds, and to implement a more physical model of how the interstellar gas cools.

\subsection{Galaxy Models and Initial Conditions}
\label{sec:ICs}

We carry out simulations of major mergers of two equal mass galaxies. We
simulate only a single galaxy mass and focus our resources on studying the
effects of feedback via AGN winds produced at small radii.  Each galaxy
model has a rotationally supported disk of gas and stars and a stellar
bulge, all embedded in a halo of dark matter.  Both the stellar and gaseous
disks have an exponential radial profile with a scale length of 3.51 kpc.
The vertical profile of the stellar disk is that of an isothermal sheet
with a scale height of 702 pc.  The vertical structure of the gas disk is
set by hydrostatic equilibrium.  The halo and stellar bulge both have
\cite{hernquist90} profiles with a halo virial and half mass radius of 229
and 102 kpc, respectively (a concentration of 9.0) and with a bulge
effective radius of 1.27 kpc.  Each galaxy has a total dynamical mass of
$1.94 \times 10^{12} \msun$, a total disk mass of $7.96 \times 10^{10}
\msun$ and a bulge of mass $2.66 \times 10^{10} \msun$; $10 \%$ of the disk
mass is gas.  We have performed an additional simulation with a disk gas
fraction of 30 \%; this run gives qualitatively similar results and so is
not discussed in detail.  The black holes are modeled as additional
collisionless particles with initial masses of $10^{5} M_\odot$.

For our fiducial simulations, each galaxy is formed from $2.4 \times 10^5$ particles:
the halo has $9 \times 10^4$ dark matter particles, the disk has $6 \times 10^4$ stellar
particles and $6 \times 10^4$ gas particles, and the bulge has $3 \times 10^4$ stellar particles.  The gravitational
force softening length is $\epsilon = 70$ pc for the disk and bulge particles 
and $\epsilon = 176$ pc for the halo particles.   We describe resolution tests in \S~3.3.

For the merger simulations, the two equal mass galaxies are placed on a
prograde orbit with an initial separation of $142.8$ kpc and an initial
velocity of each galaxy of $160$ km s$^{-1}$ directed at an angle of
$28^{\circ}$ from the line connecting the galaxies.  The corresponding
orbital energy is approximately zero.  The spin of the two galaxies are not
aligned with the orbital angular momentum of the system; the relative angle
between the spin directions is $41^{\circ}$, with one galaxy's spin making
an angle of $10^{\circ}$ degrees relative to the orbital angular momentum.

\subsection{Black Hole Accretion}
\label{sec:accretion}

In this section we briefly review the sub-grid accretion model presented in detail in DQM.  The sub-grid accretion rate on scales smaller than our resolution (both gravitational force softening and SPH smoothing) is estimated with a model 
motivated by the redistribution of angular momentum in the gas.  For gas with a surface
density $\Sigma$, a sound speed $c_s$ and a rotational angular frequency $\Omega$,
the mass accretion rate into the nuclear region is:
\begin{equation}
\label{eqn:Mdvisc}
\dot{M}_{visc} = 3 \pi \alpha \Sigma \frac{c_s^2}{\Omega}.
\end{equation}

\noindent Here $\alpha$ is a free parameter of the model characterizing not only
the  efficiency of angular momentum transport, but also the amount of gas
that turns into stars (on scales below our resolution) instead of falling into the black hole. Our fiducial values of $\alpha$ range from $0.05-0.15$, motivated (at the order of magnitude level) by comparison to the simulations of gas inflow from $\sim 0.1-100$ pc of \citet{hopkins10c}.   Over this range of $\alpha$, there is  little dependence of our simulation results on $\alpha$ (see DQM).   

To perform the sub-grid estimate of the gas properties in
equation~(\ref{eqn:Mdvisc}), we take averages of the properties of the SPH
particles inside a region of radius $R_{acc} \sim 300$ pc centered on the
black hole.  For reasons described in DQM we take this radius to be four
times the gravitational force softening, $\epsilon$, of the particles in
the simulation.

The BHs in our simulation are modeled as specially marked collisionless
particles.  {To minimize the unavoidable spurious stochastic motion of the
  BH particles due to gravitational interactions with the stellar and gas
  particles, we assign the BH particles a (fixed) tracer mass of $2\times
  10^8 M_\odot$, which is $\sim10^2$-$10^4$ times more massive than the
  other particles in the simulation.  During each simulation, we also
  compute separately the `real' mass of the BH that grows according to the
  accretion rate at small radii (with an initial mass of $10^5 M_\odot$).
  Details of this step were given in \S~2.3.2 of DQM. }

\subsection{Radiation Pressure Feedback}
\label{sec:pradfeedback}

The luminosity of the black hole is taken to be a fraction, $\eta = 0.1$, of the 
rest mass energy of the accreted material:
\begin{equation}
\label{eqn:LfromMdot}
L = {\rm min} {\left( \eta \dot{M}_{in} c^2, L_{edd} \right)}.
\end{equation}
\noindent The net accretion rate $\dot M_{\rm in}$ onto the black hole need
not equal the viscous inflow rate $\dot M_{\rm visc}$ at $\sim 100$ pc in
equation~(\ref{eqn:Mdvisc}) if there are significant outflows on scales
$\lesssim 100$ pc (roughly our resolution); we return to this point in
\S~2.4.

In the radiation pressure feedback model introduced in DQM, the accretion luminosity is coupled back into the surrounding gas by adding a total force
\begin{equation}
\label{eqn:momdep}
\dot{p}_{rad} = \tau \frac{L}{c},
\end{equation}
\noindent which is shared equally by all SPH particles inside $R_{acc}$.
The added force is directed radially away from the BH particle.  Here
$\tau$ is a free parameter of the model, representing the optical depth to
infrared (IR) radiation in the nuclear region. Our fiducial choice for $\tau$ in this work is
20.  DQM showed that $\tau \sim 20$ was required to reproduce the observed
normalization of the $M_{\rm BH}-\sigma$ correlation in their simulations.

\subsection{AGN Wind Feedback}
\label{sec:windfeedback}

In addition to the radiation pressure feedback described by
equation~(\ref{eqn:momdep}),  accretion onto the black hole can
explicitly drive a wind at radii well beneath our resolution.  Our
treatment of such a wind is motivated in particular by observations of BAL
quasars, which have $\sim 10,000 \kms$ outflows launched from near the broad
line region at $\sim 0.1$ pc (e.g., \citealt{murray1995}).  In our model,
the AGN winds at small radii carry a momentum flux given by
\begin{equation}
\label{eqn:pdwind}
\dot{p}_{w} = \tau_w \frac{L}{c},
\end{equation}
\noindent where $\tau_w$ is a further parameter of the model representing the total momentum flux in the wind.  The wind  is launched at a fixed speed $v_w$.   Thus, the rate at which mass is added to the wind is given by
\begin{equation}
\label{eqn:Mdwind}
\dot{M}_w = \tau_w \frac{L}{c v_w} = \tau_w \eta \frac{c}{v_w} \dot{M}_{in}.
\end{equation}
\noindent where we have linked the luminosity of the black hole to the
accretion rate as in equation~(\ref{eqn:LfromMdot}).  In the presence of
significant AGN winds, not all of the material entering the nuclear region,
$\dot{M}_{visc}$, actually reaches the black hole.  Instead, the net accretion
rate into the black hole, $\dot{M}_{in}$, must be reduced by the mass of
the outflow (see \citealt{ostriker10}): $\dot{M}_{in} = \dot{M}_{visc} -
\dot{M}_w$, which implies that the true black hole accretion rate is given
by
\begin{equation}
\label{eqn:Mdin}
\dot{M}_{in} = {\rm min}\left(\frac{\dot{M}_{visc}}{1 + \tau_w \eta c / v_w}, \frac{L_{edd}}{\eta c^2}\right)
\end{equation}
Note that equation~(\ref{eqn:Mdin}), not equation~(\ref{eqn:Mdvisc}),
determines the AGN luminosity and thus the magnitude of the feedback in
equations~(\ref{eqn:momdep}) and (\ref{eqn:pdwind}).

To deposit the wind momentum, we give kicks to particles inside $R_{acc}$.
Particles receiving a kick are selected stochastically with each particle
having an equal probability of being added to the wind.   The probability of a kick is chosen to ensure that the time average of the mass kicked in a given timestep $\Delta t$ is given by $\dot M_w \Delta t$.   The kick is implemented by adding a velocity $v_w$  to the current velocity of each
selected particle.  The direction of this imparted velocity is chosen to be
more heavily weighted toward the surrounding disk: if $\theta$ is the angle
between the imparted momentum and the disk normal, the probability
distribution over $x = \cos{\theta}$ is given by $p(x) = 3(1 - x^2)/4$.
For this purpose, the disk normal is defined to be the direction of the
total orbital angular momentum about the black hole of all the gas
particles inside $R_{acc}$.  This modest equatorial bias in the wind
direction is motivated by models of BAL quasars \citep{murray1995}.
Nonetheless, this choice is not that critical:  the efficacy of the feedback is largely determined by the outflow momentum/energy flux that is directed within the solid angle subtended by the surrounding ISM.  For our fiducial model, $\sim 40 \%$ of the momentum flux is directed within one scale-height of the disk at $\sim 100$ pc.  Models with isotropic kicks require slightly large values of $\tau_w$ to give results similar to our fiducial model because less of the feedback is directed towards the surrounding ISM.

{Despite the similarity between  equations~(\ref{eqn:momdep}) and (\ref{eqn:pdwind}), our two feedback processes affect the gas particles within $R_{acc}$ in different ways.
  In the case of our simple radiation pressure model, all gas particles within $R_{acc}$
  receive an additional acceleration set by equation~(\ref{eqn:momdep})
   and are pushed outward from the BH.  As shown in DQM, this feedback process evacuates the central gas  reservoir and lowers the BH accretion rate and the final BH mass, but the velocities of the gas particles are not so large as to produce significant large-scale galactic outflows.    In
  our AGN wind model, by contrast, a small fraction ($\la 5$\%) of gas
  particles within $R_{acc}$ receives an instantaneous large wind velocity
  of 3000 to $10,000 \kms$.  These particles leave $R_{acc}$ within $ \la
  10^5$ years and, as we show below, can drive a galaxy-scale outflow via their interaction with the ambient ISM.}

Our implementation of wind feedback adds two additional input parameters to
the simulation.  The first, $v_w$, describes the launch speed of the winds
while the second, $\tau_w$, describes the total momentum flux in the wind.
Our default wind speed is motivated in part by observations of BAL quasars
and theoretical models of the origin of such winds via line driving in the
accretion disk at $\sim 0.1$ pc (though our model should not be taken as a
literal implementation of this physics).  Observed wind velocities are
$\sim 10,000 \kms$.  Models of line driving lead to momentum fluxes in the
wind of $\lesssim L/c$, i.e., $\tau_w \lesssim 1$ because the lines do not
typically completely cover the continuum \citep{murray1995}.  As we show below, however, larger values of $\tau_w$ are required for AGN outflows to have a significant effect on the gas dynamics and star formation history in galactic nuclei.

Observationally, momentum fluxes are difficult to infer in most cases
because of ambiguities in the absolute density/radial scale at which the
absorption occurs.  In a handful of low-ionization BALs (in particular,
FeLoBALs), this degeneracy has been broken, implying momentum fluxes $\sim
0.3-5 L/c$, i.e., $\tau_w \sim 0.3-5$ (with significant uncertainties; see \citealt{moe09, bautista10,dunn10,claude11}).  It
remains uncertain whether these values of $\tau_w$ are representative of
the entire BAL-quasar population.  We take $\tau_w \simeq 5$ as our
fiducial value but also explore the range  $1\le \tau_w \le 10$.  Our fiducial value of $\tau_w = 5$ was chosen for three reasons that will become clear later in the paper:  (1) the AGN wind then leads to a galaxy-scale outflow that significantly influences the surrounding ISM dynamics, (2) the final BH mass in the simulations is reasonably consistent with the observed $M_{\rm BH}-\sigma$ relation for $\tau_w \simeq 5-10$.  (3)  Observations of high speed atomic and molecular outflows in local ULIRGs suggest $\tau_w \sim 10$ \citep{feruglio10,chung11,rupke11,sturm11}.    We discuss physical mechanisms that can produce such powerful AGN winds in \S \ref{sec:discussion}.

\subsection{ISM Model and Gas Cooling}
\label{sec:cooling}

In the subgrid ISM model of SH03, gas with densities above the star formation threshold approaches an effective thermal energy, $u_{eff}$, set by a balance between  cooling and the feedback from star formation.  If processes such as shocks or adiabatic expansion/compression cause the internal energy to deviate from $u_{eff}$, the differences decay on the timescale given by eq.~(12) of SH03.  This decay timescale is set by the subgrid model rather than the true cooling time of the gas.  

One consequence of SH03's ISM model is that sufficiently dense shock heated gas does not cool on its cooling timescale.   
In this paper we show that AGN winds can shock heat gas in the ISM to above the escape speed -- this contributes to driving a galactic wind.   To ensure that the cooling of the shock heated gas is correct, we modified the ISM model of SH03 to better match the expected cooling rate for gas with roughly a solar metallicity.  To compute the local cooling rate we use a fit to  \cite{sutherland93} for 
gas with solar metallicity. The fit is essentially eq.~(12) of \cite{sharma10} but with the cooling rate increased by a factor of two when the temperature is between $\sim 3 \times 10^4$ K and $3 \times 10^7$ K (to account for solar metallicity).  
From this cooling rate we compute a cooling time, $t_{cool}$; differences between the thermal energy and the subgrid $u_{eff}$ decay on  $t_{relax} = t_{cool}$.   For most of the relevant range of temperature and density in our simulations, this cooling time is shorter than the relaxation 
time of SH03.  

\begin{table*}
  \caption{Simulation Parameters}
  \label{tab:simparm}
  \begin{tabular}{lcccccccccl}
   \hline
   Run & $\tau_w$ & $v_w$ & ISM & $M_{BH,f}$ & $M_{*,f}$ & $M_{*,p}$ & $f_{\rm disk}$ & $f_{\rm out}$ & $M_{\rm out}$ & Notes \\
   Name & & [$\kms$] & Model & [$10^7 \msun$] & [$10^9 \msun$] & [$10^9 \msun$] & & & [$10^9 \msun$] & \\
   \hline
   wo1-10  & 1  & 10000 & a & 56.5 & 8.67 & 1.79 & 0.014 & 0.343 & 5.47 & wo = wind only \\
   wo3-10  & 3  & 10000 & a & 17.9 & 7.35 & 1.05 & 0.002 & 0.472 & 7.52 & \\
   wo5-10  & 5  & 10000 & a & 12.2 & 7.36 & 1.02 & 0.002 & 0.477 & 7.60 & \\
   wo5-7   & 5  & 7000  & a & 5.36 & 10.7 & 2.91 & 0.041 & 0.175 & 2.78 & \\
   wo5-3   & 5  & 3000  & a & 6.65 & 12.1 & 5.23 & 0.059 & 0.056 & 0.89 & \\
   wo10-10 & 10 & 10000 & a & 6.20 & 7.35 & 0.92 & 0.002 & 0.466 & 7.42 & \\
   wo10-7  & 10 & 7000  & a & 2.99 & 10.8 & 2.55 & 0.014 & 0.191 & 3.04 & \\
   wo10-3  & 10 & 3000  & a & 3.96 & 12.2 & 4.70 & 0.069 & 0.060 & 0.96 & \\
   \hline
   wp5-10   & 5  & 10000 & a       & 4.38 & 10.9 & 3.11 & 0.006 & 0.196 & 3.13 & fiducial sim; wp = wind + prad\\
   wp5-7    & 5  & 7000  & a       & 5.04 & 10.9 & 3.77 & 0.016 & 0.175 & 2.79 & \\
   wp5-3    & 5  & 3000  & a       & 2.63 & 12.8 & 4.02 & 0.159 & 0.102 & 0.36 &  \\
   wp1-10   & 1  & 10000 & a       & 11.8 & 12.4 & 3.86 & 0.006 & 0.100 & 1.61 & \\
   wp3-10   & 3  & 10000 & a       & 6.15 & 11.9 & 4.30 & 0.006 & 0.149 & 2.37 &  \\
   wp10-10  & 10 & 10000 & a       & 3.36 & 10.3 & 3.21 & 0.002 & 0.268 & 4.26 &  \\
   wpfg30   & 5  & 10000 & a       & 10.2 & 39.5 & 4.54 & 0.003 & 0.148 & 7.06 & $f_{g,i} = 0.3$ \\
   wpsh     & 5  & 10000 & c       & 6.14 & 10.8 & 3.41 & 0.002 & 0.234 & 3.73 &  \\
   wpcomp   & 5  & 10000 & b       & 6.30 & 10.8 & 3.21 & 0.004 & 0.276 & 4.40 &  \\
   wpcompl  & 5  & 10000 & b$^{*}$ & 5.01 & 11.4 & 3.96 & 0.007 & 0.181 & 2.88 &  \\
   wpinst   & 5  & 10000 & d       & 5.90 & 10.4 & 2.67 & 0.018 & 0.216 & 3.44 &  \\
   wplr     & 5  & 10000 & a       & 5.28 & 10.2 & 4.45 & 0.004 & 0.307 & 4.89 & Lower resolution \\
   wphr     & 5  & 10000 & a       & 4.77 & 11.5 & 3.94 & 0.012 & 0.166 & 2.65 & Higher resolution \\
   po       & 0  & 0  & a       & 9.33 & 13.6 & 4.92 & 0.051 & 0.007 & 0.11 & No wind feedback \\
   pofg30   & 0  & 0  & a       & 25.9 & 44.5 & 6.45 & 0.028 & 0.007 & 0.32 & $f_{g,i} = 0.3$, No wind feedback \\
   \hline
  \end{tabular}

  \medskip
  Model parameters and key quantities in our galaxy merger simulations.
  The top section lists runs with only the AGN wind feedback model (parametrized by the wind speed $v_w$ and momentum flux $\tau_w$).  The bottom
  section lists runs with both the AGN wind model and the radiation pressure model with
  $\tau = 20$.   ISM model refers to one of four following options:
  (a) our fiducial cooling model (see \S \ref{sec:cooling}), (b) same as (a) but including Compton heating with
  $T_C = 2 \times 10^7$ K, (b$^{*}$) same as (a) but including Compton heating with
  $T_C = 10^6$ K, (c) same as in SH03, and (d) the gas returns to the effective EOS of
  SH03 instantaneously.  Other columns list the key properties of the black holes, stars, and gas
  in the simulations:  $M_{BH, f}$ is the final black hole mass,  $M_{*, f}$ is the total
mass of new stars formed by the end of the simulation,  $M_{*,p}$ is
 the total mass of new stars formed after the peak of star formation,
 $f_{\rm disk}$ is the fraction of the gas mass within
3 kpc of the BH at the end of the simulation, $f_{\rm out}$
is the fraction of the gas mass at large
distances ($\gtrsim 10$ kpc) above the orbital plane at the end of the
simulation, and $M_{\rm out}$ is the corresponding gas mass.   $f_{\rm disk}$ is a proxy for how much gas remains in the galaxy at late times while $f_{\rm out}$ is a proxy for the fraction of the gas that is in the outflow at late times; both $f_{\rm disk}$ and $f_{\rm out}$ are normalized to the
  initial total gas mass of $1.59\times 10^{10} M_\odot$.  
\end{table*}

In addition to the above modification to the cooling rate, we also consider the role of inverse Compton cooling/heating.   To do so, we modify the cooling/relaxation timescale of the gas to be
\begin{equation}
\label{eqn:trelcomp}
t^{-1}_{relax} = t^{-1}_{cool} + t^{-1}_C.
\end{equation}
\noindent where the (non-relativistic) Compton time is given by
\begin{equation}
\label{eqn:tcomp}
t_C = 10^7 {\rm yr} \left( \frac{R}{300 {\rm~pc}} \right)^2 \left( \frac{L}{10^{46} {\rm~erg} {\rm~s}^{-1}} \right)^{-1},
\end{equation}
with $R$  the distance to the BH and $L$  the associated AGN 
luminosity.  We do not consider any radiative transfer effects in this paper, which in reality can modify both the luminosity $L$ seen at a given radius and the local Compton temperature at a given radius.    Instead, we take the Compton temperature to be that appropriate for the mean spectrum of luminous AGN, including the effects of obscuration:  $T_C \simeq 2 \times 10^7$ K \citep{sazonov04}.

In the limit that the Compton timescale is short compared to the two-body cooling time, we no longer relax the thermal energy of the gas to the effective equation of state value for the energy, $u_{eff}$, but rather to $u_C = 3/2 k T_C$.   To transition between these two limits, we in general let the thermal energy of gas relax to:
\begin{equation}
\label{eqn:urelaxto}
u_{relax} = \frac{t_{cool} u_C + t_C u_{eff}}{t_C + t_{cool}}.
\end{equation}

For dense gas, atomic cooling dominates and the gas rather quickly approaches the sound speed associated with the effective equation of state.  For gas densities characteristic of the ISM in the central kpc of our model galaxies ($\sim 10-10^3$ cm$^{-3}$) the sound speeds are $\sim 40 \kms$ and thus the gas is primarily rotationally supported rather than pressure supported.  This justifies our use of the viscous accretion rate in equation \ref{eqn:Mdvisc}.

\section{Results}
\label{sec:results}

Table~\ref{tab:simparm} lists the parameters of the galaxy merger
simulations presented in this work.  We consider a single galaxy mass 
and merger orbit (see \S~\ref{sec:ICs}), and explore the effects
of including AGN wind feedback, both with and without radiation pressure.
For the AGN winds, we vary both the total momentum flux in the wind
($\tau_w$) and the wind speed $v_w$, with fiducial values of $\tau_w = 5$
and $v_w = 10,000 \kms$.  

In varying the AGN wind parameters, we are not guaranteed that the resulting BH mass  will be consistent with the $M_{BH}-\sigma$ relation.   In such cases, the effects of the AGN wind might not be realistic because of the unphysical BH mass.   To quantify this, we present results with and without the radiation pressure feedback model explored in DQM.   Our simple model of radiation pressure feedback produces model galaxies roughly on the observed $M_{\rm BH}-\sigma$
correlation for $\tau \sim 20$, the value used here (DQM).   For the present purposes these calculations are useful primarily because they allow us to study the effects of AGN wind feedback for systems in which the BH is guaranteed to be approximately on the $M_{BH}-\sigma$ relation.   We also separately carry out simulations with wind feedback alone.

\subsection{Effects of AGN Winds}
\label{sec:fidsim}

\subsubsection{Black hole accretion and star formation rates}

\begin{figure}
\includegraphics[width=84mm]{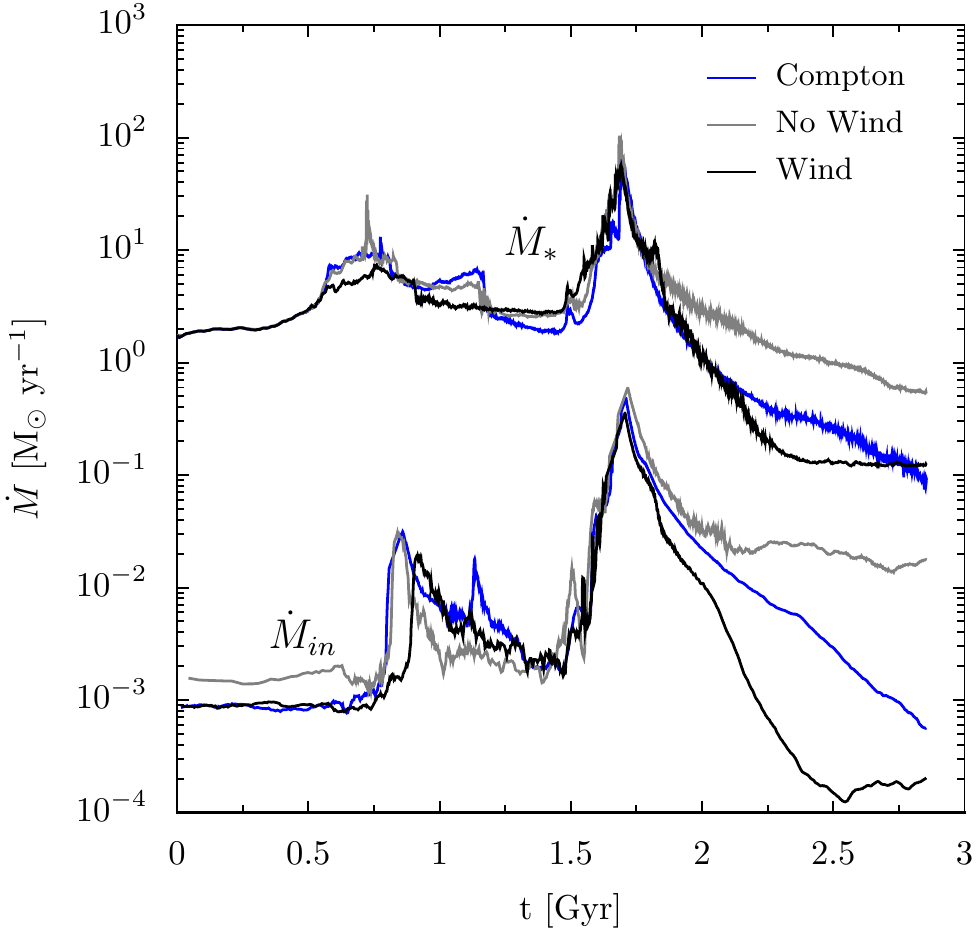}
\caption{ The star formation rate, $\dot{M}_{*}$, and accretion rate into
  the black hole, $\dot{M}_{in}$, as a function of time for the fiducial
  simulation with both AGN wind and radiation pressure feedback (run
  `wp5-10' in Table~1; black), the run with only the radiation pressure
  feedback (run `po'; gray), and the run that included Compton heating
  (run `wpcomp'; blue).  }
\label{fig:mdotfid}
\end{figure}

\begin{figure}
\includegraphics[width=84mm]{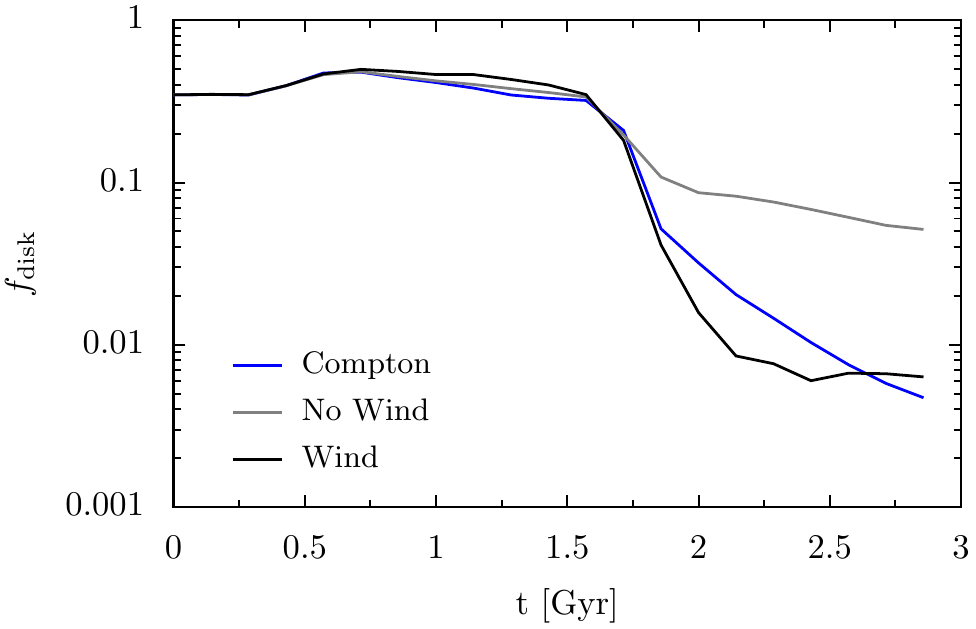}
\caption{ The fraction of the total gas mass inside a sphere of radius 3 kpc
  (relative to the total {\it initial} gas mass) centered on the black hole for
  the same three runs shown in Fig.~1.
   When the AGN wind model is included, the central material is
  rapidly ejected after final coalescence ($t \sim 1.7$ Gyr).  }
\label{fig:m3kpc}
\end{figure}

\begin{figure}
\includegraphics[width=42mm]{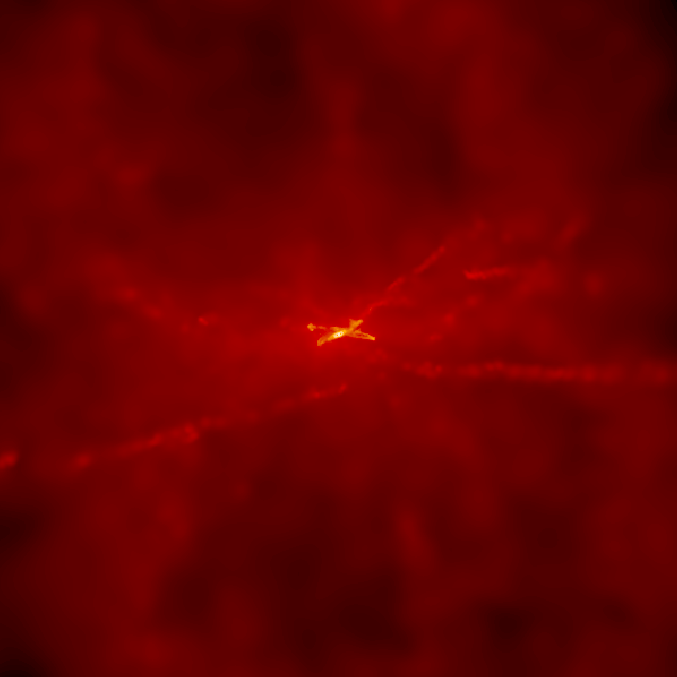}\includegraphics[width=42mm]{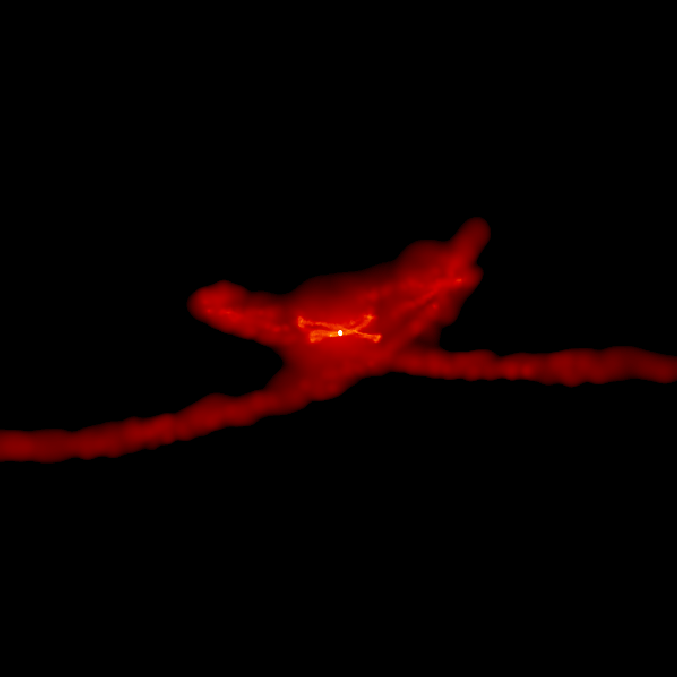}
\caption{The projected gas density for the fiducial simulation (left) and the run without
AGN wind feedback (right) at a time just after the final merger ($t=1.71$ Gyr).  
Brighter color corresponds to higher density. The images are edge-on to the
plane of the orbit and the box size is 280 kpc. While both simulations have 
material at large distance in tidal tails, there is significantly more material
out of the orbital plane with AGN wind feedback.  These images were generated
using SPLASH \citep{price07}.
}
\label{fig:densplot}
\end{figure}

Figure \ref{fig:mdotfid} shows the net accretion rate into the black hole,
$\dot{M}_{in}$, and the star formation rate, $\dot{M}_{*}$, as a function
of time (summed over both galaxies) for three different simulations: the
fiducial simulation (black curve) with radiation pressure ($\tau=20$) and
AGN wind feedback ($\tau_w = 5$ and $v_w = 10,000 \kms$), a run identical
to the fiducial run but without the wind feedback (gray), and a run with
both wind and radiation pressure feedback that includes Compton
heating/cooling in the thermodynamics of the gas (blue).  We first describe
the effects of the AGN wind and later return to the role of Compton
heating/cooling.  

The inclusion of the wind feedback has little effect on either the
accretion or star formation rate before the final coalescence of the two
galaxies at $t \sim 1.7$ Gyr.  This is {\em a priori} surprising because
equation~(\ref{eqn:Mdin}) implies that for a given set of conditions at
large radii (that determines $\dot M_{\rm visc}$) the BH accretion rate at
small radii is a factor of $1+ \tau_w\eta c/v_w =16$ smaller for the
simulation with AGN winds than for the run with just radiation pressure.
After a small number of time steps, however, the feedback due to radiation
pressure is so effective in all of the simulations in Figure
\ref{fig:mdotfid} that the physical conditions at small radii quickly
adjust so that there is a balance between radiation pressure and gravity.
This sets the BH accretion rate $\dot M_{in}$ to be $\sim f_g
\sigma^4/(\tau \eta c G)$ \citep{debuhr10} independent of the presence of
the AGN wind removing mass from the nuclear region (where $f_g$ and
$\sigma$ are the gas fraction and velocity dispersion at small radii,
respectively).

Although the AGN wind feedback has little effect on the early time star
formation and BH accretion, Figure \ref{fig:mdotfid} shows that after final
coalescence both the BH accretion rate and the star formation rate decrease
significantly more rapidly with the inclusion of the AGN wind.  The effects
of the BHs feedback are maximized at and after final coalescence because
this is when the BH reaches its final mass (to within a factor of a few)
and when the BH accretion rate is largest.  Thus, the dynamics at and after
final coalescence is the most sensitive to the details of the feedback
physics.

To assess why the late-time accretion and star formation are suppressed by the inclusion of the AGN wind, Figure \ref{fig:m3kpc} shows the gas mass fraction (normalized to the gas mass at the start of the simulation) within 3 kpc of the BH
as a function of time, for the same three simulations as in Figure \ref{fig:mdotfid}.\footnote{The precise choice of 3 kpc is somewhat arbitrary but the results in Figure \ref{fig:m3kpc} are relatively robust to changes in this choice.}  Tidal interactions drive gas into the center of the two galaxies (or the merged remnant) after the first close passage ($t \sim 0.5$ Gyr) and at final coalescence ($t \sim 1.7$ Gyr), maintaining a large nuclear gas fraction in spite of the intense star formation.
The resulting increase in the BH accretion rate (seen also in Fig.~\ref{fig:mdotfid}) significantly increases the strength of the AGN feedback.  In the simulation including AGN winds, this feedback efficiently removes material from the central regions of the galaxy, reducing the amount of gas inside the central 3 kpc of the merged system by a factor of $\sim 10$ relative to the simulation without explicit AGN winds.

Figure \ref{fig:densplot} shows images of the projected gas density for the fiducial simulation ({\em left}) and the corresponding simulation without the AGN wind ({\em right}) at $t=1.71$ Gyr, shortly after  final coalescence.  The images are 280 kpc on a side and show a roughly edge-on view of the orbital plane of the two galaxies.
While there is material at large distances in the simulation without the
AGN wind, these features are the tidal tails generated during the merger
and most of this material is near the orbital plane.  Thus our simple
implementation of radiation pressure feedback is not efficient at unbinding
gas from the galaxy, though it is very effective at regulating the growth
of the BH itself (DQM).  By contrast, in the simulation with the AGN wind,
Figure \ref{fig:densplot} (left panel) shows that there is significantly more material
blown out of the galaxy, especially in the directions perpendicular to the
orbital plane.  Quantitatively, at the end of the simulation the mass of
gas at large distances ($|z| > 10$ kpc) from the orbital plane is $\simeq 3
\times 10^9 \msun$ in the simulation including the AGN wind, about 20 times
larger than in the simulation with only radiation pressure feedback (see
Table~\ref{tab:simparm} and Fig.~\ref{fig:woblowout} below).  A corollary
of this efficient removal of gas by the AGN wind is that the total stellar
mass formed during the simulation is $\sim 20 \%$ smaller in the case with
the AGN wind; most of this suppression in star formation happens at late
times, after the final coalescence of the two galaxies (see
Fig.~\ref{fig:mdotfid}).

\begin{figure}
\includegraphics[width=84mm]{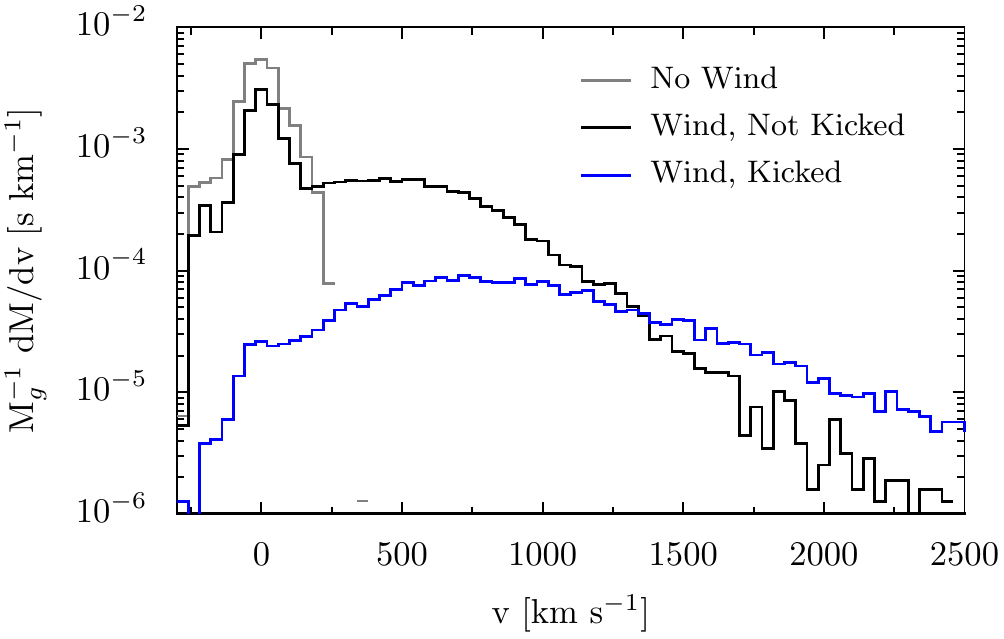}
\caption{Effects of AGN winds on the distribution of radial velocities of
  the gas particles (relative to the black hole) at the end of the
  simulation.  Few gas particles have velocities greater than $\sim 200
  \kms$ in the run with only radiation pressure (run `po'; grey).  By contrast, in the
  fiducial simulation with the AGN wind,  a significant fraction of the particles
  have velocities above a few 100 $\kms$.  About 25\% of these outflowing particles
  were explicitly added to the wind and kicked with an initial wind velocity of 10,000 $\kms$ (blue),
  while the rest were not explicitly added to the wind (black) initially but were accelerated
  due to hydrodynamic interactions with the `kicked' high-speed particles.}
\label{fig:vpdf}
\end{figure}

\begin{figure}
\includegraphics[width=84mm]{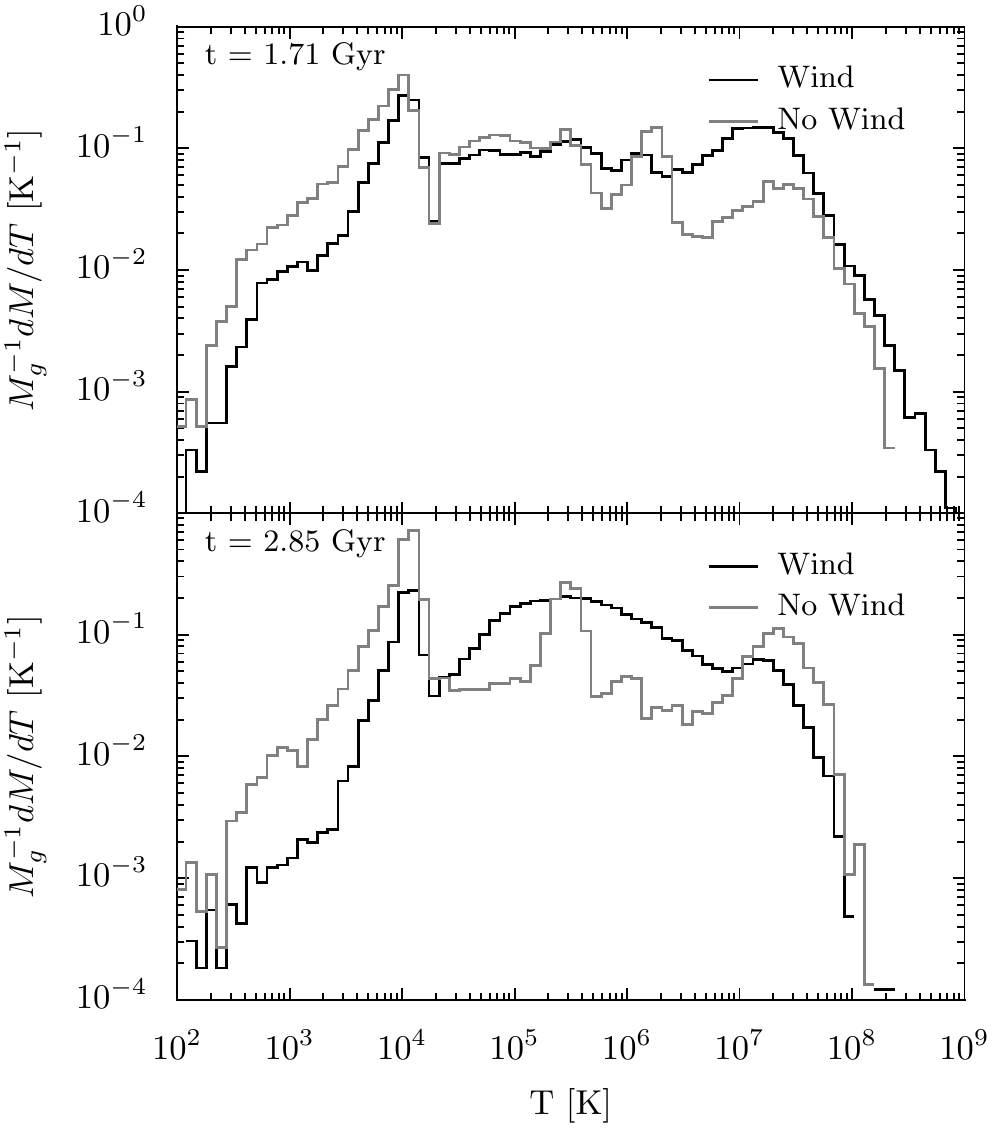}
\caption{Effects of AGN winds on the  mass weighted distribution of gas
  temperatures.  The same two simulations as in Fig.~4 are shown at two
  times: near the peak of accretion at 1.71 Gyr ({\em top}), and at the end of the
  simulation at 2.85 Gyr ({\em bottom}).  The temperature distributions are 
  normalized to have a unit area.  The excess of gas at high temperatures near the peak of BH accretion ({\em top}) in the case with wind feedback is due to the shock heating of the ISM gas.  }
\label{fig:Tpdf}
\end{figure}

\subsubsection{Impact on the ISM}

The evacuation of the central part of the galaxy by the AGN wind would not
be surprising if the majority of the material ejected was explicitly added
to the wind by our deposition of momentum (i.e., if the unbound mass was
primarily material that was explicitly 'kicked').  We find that this is not
the case.  At the end of the fiducial simulation ($t = 2.85$ Gyr), of the
33,028 gas particles at large transverse position {($|z| > 10$ kpc)}, only 8,904 of the
particles have been explicitly kicked.  The other 72\% (82\% by mass) have
been ejected because of hydrodynamic interactions with wind material.  In
addition, at any time only a small fraction $\sim 5 \%$ of the gas inside
$R_{acc}$ (the accretion/feedback region) has been explicitly added to the
wind.  These results demonstrate that the majority of the galactic outflow
arises self-consistently due to the interaction between the AGN wind that
we initialize and the surrounding ISM.  

Figure \ref{fig:vpdf} illustrates this point in a different way: here we
show the radial velocities (relative to the BH) of the gas particles at the
end of the simulation, without (gray) and with (black and blue) wind
feedback.  Those particles that \emph{have not} been explicitly kicked and
added to the wind are shown in black while those that have been explicitly
kicked are shown in blue.  In the case with the AGN wind, there is a
significant fraction of the gas with velocities $\sim 300-1000 \kms$,
larger than {\em any} of the gas in the simulation without the wind
feedback (gray).  Moreover, the outflow velocities are similar for
particles that have, and have not, been explicitly kicked.  The particles
that are explicitly kicked receive initial impulses of $v_w = 10,000 \kms$
and yet have final speeds of a few thousand $\kms$.  This result indicates
that the wind we initialize sweeps up a significant amount of material in
the ISM of the surrounding galaxy, driving additional gas out of the galaxy
at high velocities $\sim 300-1000 \kms$.

Figure~\ref{fig:Tpdf} shows the mass weighted temperature distribution (normalized to the total gas mass) near the peak of accretion ($t = 1.71$ Gyr; {\em top}) and at the end of the simulation (t = 2.85 Gyr; {\em bottom}) for the run with AGN wind feedback (black) and without (gray).   The presence of the AGN wind leads to a significant amount of mass heated to above $\sim 10^7$ K when the BH is near its peak accretion rate  ({\em top} panel in Fig.~\ref{fig:Tpdf}).   This is due to gas that has been shock heated by the AGN wind.   Much of this shock heated gas is ultimately able to escape the host galaxy -- this gas expands outwards and cools adiabatically, leading to the large excess of $\sim 10^{5-6}$ K gas at the end of the simulation with the AGN wind  ({\em bottom}).  

In our calculations that include a simple treatment of inverse Compton scattering off of the AGN's radiation field, we take the Compton temperature to be $T_C = 2 \times 10^7$ K,  the average observed value \citep{sazonov04}.    This  is similar to the peak in the temperature distribution we find in simulations without Compton scattering ({\em top} panel of Fig.~\ref{fig:Tpdf}).  As a result Compton scattering mildly suppresses the high temperature tail of the gas shock heated by the AGN wind.   This in turn suppresses the amount of mass blown out of the inner regions of the galaxy at late times (see Fig.~\ref{fig:m3kpc}).   Nonetheless, there is still a factor of $\sim 3-10$ times less gas in the inner 3 kpc in the simulation with the AGN wind and Compton cooling relative to the simulation without the AGN wind.  In addition, the star formation rate and BH accretion rate are significantly lower at late times even when Compton cooling is taken into account (Fig.~\ref{fig:mdotfid}). 

\subsection{Dependence on Feedback Model Parameters}
\label{sec:vw}

In the previous section we demonstrated that an AGN wind generated at small radii can produce a  galaxy-wide outflow that unbinds several $10^9 \, M_\odot$ of gas and suppresses star formation in its host galaxy.   In this section we explore in more detail how this phenomenon depends on the properties of the AGN wind.

To quantify the effects of the AGN wind, we calculate the following
quantities in all of our simulations and list the results in
Table~\ref{tab:simparm}: (1) the final BH mass $M_{BH, f}$, (2) the total
mass of new stars formed by the end of the simulation, $M_{*, f}$, (3) the
total mass of new stars formed after the peak of star
formation,\footnote{We distinguish between $M_{*,p}$ and $M_{*,f}$ because
  Figure \ref{fig:mdotfid} shows that the AGN wind's largest effect on the
  star formation history is at late times, after the peak of BH accretion
  and star formation.}  $M_{*,p}$, (4) the fraction of the gas mass within
3 kpc of the BH at the end of the simulation, $f_{\rm disk} = M_{g,r<3
  \rm{kpc}}/M_{g, tot}$, and (5) the fraction of the gas mass at large
distances ($\gtrsim 10$ kpc) above the orbital plane at the end of the
simulation, $f_{\rm out} = M_{g,|z|>10 \rm{kpc}}/M_{g, tot}$.  {The
  fractions $f_{\rm disk}$ and $f_{\rm out}$ are both normalized to the
  {\it initial} total gas mass of $1.59\times 10^{10} M_\odot$ (from both
  galaxies);} $f_{\rm disk}$ is a proxy for how much gas remains in the
galaxy while $f_{\rm out}$ is a proxy for the fraction of the gas that is
in the outflow at late times.  Table \ref{tab:simparm} also quotes $M_{\rm
  out}$, the absolute mass in the outflow at the end of the simulation.

\begin{figure*}
\includegraphics[width=170mm]{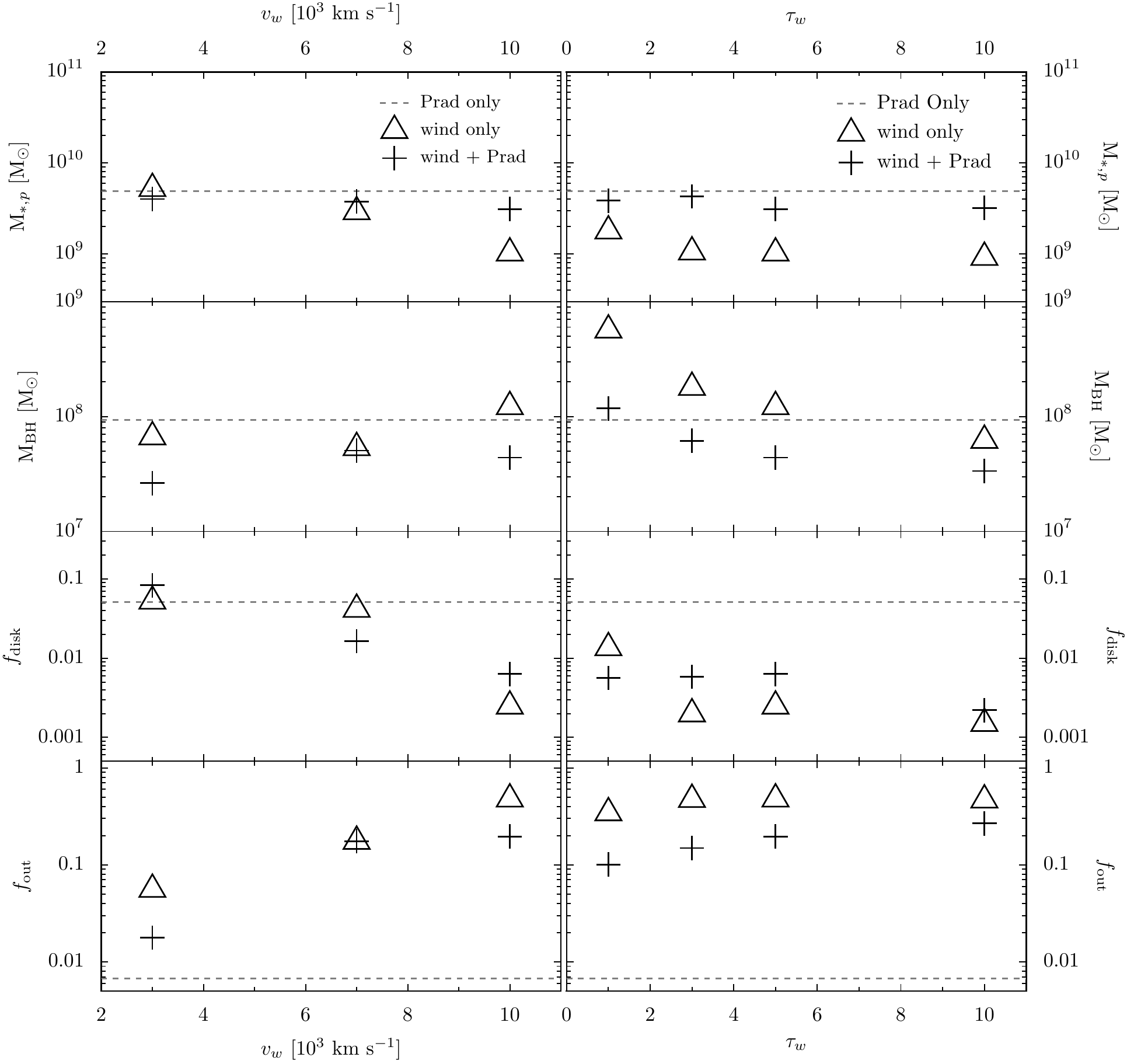}
\caption{ Key quantities characterizing the effects of the AGN wind model
  as a function of wind speed $v_w$ (left column) and momentum flux
  $\tau_w$ (right column); see Table~1 for more details.  These results quantify describe how the  properties of the galaxy merger remnant at the end of the simulation depend on the parameters characterizing the AGN wind at small radii.    In each panel, the triangles label the runs with only the AGN wind feedback, the crosses label the runs with both AGN wind and
  radiation pressure feedback, and the dashed line is for the simulation
  with only radiation pressure feedback.  The four rows (from top to down)
  show: the mass $M_{*,p}$ of new stars formed after the peak of star
  formation during the simulation; the final black hole mass $M_{BH,f}$;
  the mass fraction $f_{\rm disk}$ of gas at the end of the
  simulation within 3 kpc of the black hole, and the gas mass fraction $f_{\rm out}$ at the end of the simulation that is at large distances from the orbital plane: $|z| > 10$ kpc .  Both $f_{\rm disk}$ and $f_{\rm out}$ are normalized to the total initial gas mass in the system.}
\label{fig:woblowout}
\end{figure*}

Figure~\ref{fig:woblowout} compares these key quantities for simulations
utilizing only the AGN wind feedback model (triangles) and for simulations with both
radiation pressure and AGN wind (crosses).  For comparison, we also show the
results of the simulation with only radiation pressure feedback, taking $\tau
= 20$ (dotted lines).  In the {\em left} set of panels, we fix $\tau_w = 5$
and vary the wind speed from $3,000-10,000 \kms$.  In the {\em right} set
of panels, we fix $v_w = 10,000 \kms$ and vary $\tau_w$ from $1-10$.
   
Figure~\ref{fig:woblowout} ({\em left} panel) shows that the final BH mass
is relatively independent of the speed of the AGN wind for a given value of
$\tau_w$.  By contrast, for simulations with the AGN wind model alone, the
{\em right} panel shows that the final BH mass decreases by a factor of
$\sim 10$ as $\tau_w$ increases from $1-10$.\footnote{The same effect is
  much less evident in the simulations with radiation pressure and the AGN
  wind (crosses in the {\em right} panels) because the former largely sets
  the BH mass.}
This is consistent with the analytic results of \citet{king03} and \citet{murray05}, and the numerical results of DQM, in which the final BH mass decreases linearly with the total momentum supplied
by the AGN, be it in the form of a wind or radiation.  In particular, when the accretion rate is Eddington limited, the net force associated with the feedback is $\propto \tau_w L_{edd} \propto \tau_w M_{BH}$ (eq. \ref{eqn:pdwind}).   Thus the BH mass at which feedback is able to overcome the gravity of the gas in the galactic nuclei scales as $M_{BH} \propto \tau_w^{-1}$.  

Although the BH mass is only a weak function of $v_w$ at fixed $\tau_w$, the properties of the gas at the end of the simulation depend strongly on $v_w$ ({bottom, left} panels in Fig.~\ref{fig:woblowout}).   Specifically, a larger wind speed $v_w$ at fixed $\tau_w$ leads to more efficient blow out of gas from the galaxy at late times (smaller $f_{\rm disk}$ and larger $f_{\rm out}$).   For a given value of $\tau_w$, the momentum flux in the outflow ($\dot p_w = \tau_w L/c$; eq.~\ref{eqn:pdwind}) is independent of $v_w$.   By contrast, the energy flux in the outflow ($\dot E_w = \dot p_w v_w$) increases for a larger wind speed.   Thus a larger wind speed leads to more efficient shock heating of the gas in the galactic nucleus. 
A larger value of $\tau_w$ has a similar effect, increasing both the momentum and energy fluxes in the wind, and producing a more powerful galaxy-scale outflow (bottom, right panels in Fig.~\ref{fig:woblowout}).   

Overall, then, the galaxy-wide outflow is a consequence of both the wind launched from small radii sweeping up and driving out ambient gas and the wind generating a shock that heats and unbinds ambient gas.    This is why the gas content in the outflow -- as measured by $f_{\rm out}$ in Figure~\ref{fig:woblowout} --  increases with both increasing $v_w$ and increasing $\tau_w$.    In general the simulations with AGN wind feedback alone have somewhat higher outflow rates than the simulations with AGN winds and radiation pressure.   The reason is that the feedback is then provided entirely by the wind, enabling it to be more effective and unbind a larger amount of gas.

The mass of new stars formed in our simulations decreases for larger $v_w$ and/or $\tau_w$ (top panels in Fig.~\ref{fig:woblowout}).  This is a simple consequence of the galaxy-scale outflow being more powerful for larger $v_w$ and/or $\tau_w$.   The net effect of the outflow on the  stellar mass formed in the simulation is, however, relatively modest (typically $\sim 10-20 \%$).  The AGN wind's most prominent effect on the star formation history is that it can suppress the late-time star formation, as highlighted in Figure~\ref{fig:mdotfid}.   It is also important to note, however, that the outflow does not always have this effect.   For example, the top left panel of Figure \ref{fig:woblowout} shows that simulations with outflows having $\tau_w = 5$ and $v_w = 3000 \kms$ form a very similar amount of new stars (both in total and after the peak of star formation) to simulations with the radiation pressure feedback model alone.   This is consistent with the fact that for these parameters, the outflow does not appreciably change the gas content in the galactic disk ($f_{\rm disk}$ in Fig.~\ref{fig:woblowout}), although it does unbind a modest amount of additional gas $\sim 3 \times 10^8 \msun$  (see $f_{\rm out}$ in Fig.~\ref{fig:woblowout} and $M_{out}$ in Table \ref{tab:simparm}).   

\subsection{Numerical Tests}
\label{sec:lastresults}

We argued in \S \ref{sec:fidsim} and \ref{sec:vw} that the galactic outflow seen in our simulations is a consequence of both the AGN wind sweeping up ambient gas (momentum conservation) and the AGN wind shock heating ambient gas and unbinding it (energy conservation).   The former mechanism does not depend sensitively on the thermodynamics of the gas, while the latter does.   As described in \S \ref{sec:cooling} we modified the ISM equation of state of SH03 in order to include two-body cooling at solar metallicity; we also included a simple treatment of inverse Compton scattering off of the AGN's radiation field in some of our simulations (see Fig. \ref{fig:mdotfid}).  

To further assess the sensitivity of our results to the thermodynamics of
the gas, we carried out a number of additional tests.  In particular, two
extremes are to use the SH03 model as is (in which gas that deviates
from the equilibrium equation of state relaxes back on a timescale
different from, and typically longer than, the true cooling time), or
to force gas to return to the effective equation of state of SH03
instantaneously (thus removing the possibility of strong shock heating
unbinding gas).  The results of these calculations with our standard AGN
wind parameters are given in Table \ref{tab:simparm}, labeled wpsh and
wpinst, respectively.  In both cases, the AGN wind drives a galaxy-scale
outflow with an ejected mass $\sim 3 \times 10^9 \msun$, very similar to
our fiducial simulation.  However, these 'extreme' ISM models have effects
that are not surprising: SH03's equation of state (in which the gas
effectively cools more slowly) is more effective at suppressing star
formation and clearing gas out of the nucleus (wpsh), while the simulation
with instantaneous cooling (wpinst) retains a larger nuclear gas disk at
the end.  Overall, however, these differences are  modest -- they are
smaller, e.g., than the uncertainties introduced by the relatively poor
current constraints on the total momentum/energy flux in AGN winds ($\tau_w$).
This test demonstrates that our treatment of the ISM thermodynamics is probably
sufficiently accurate to assess the impact of AGN winds at the order of
magnitude level.  However, future calculations that incorporate a more
realistic multi-phase, turbulent ISM (e.g., \citealt{hopkins11}) will be
very valuable.

Recently, \cite{durier11arxiv} showed that after injection of energy, either thermally
or kinetically, inaccuracies can arise in the subsequent evolution of the
SPH particles when each is assigned a separate timestep, as occurs in GADGET.  To
ascertain if our feedback is subject to the innacuracies they report, we
reran the fiducial simulation with an implementation of their timestep limiter.  
All of our key qualitative conclusions  are unchanged, though there is a modest ($\sim 30$ \%) change in the post merger star formation and disk fraction ($f_{\rm disk}$).

\begin{figure}
\includegraphics[width=84mm]{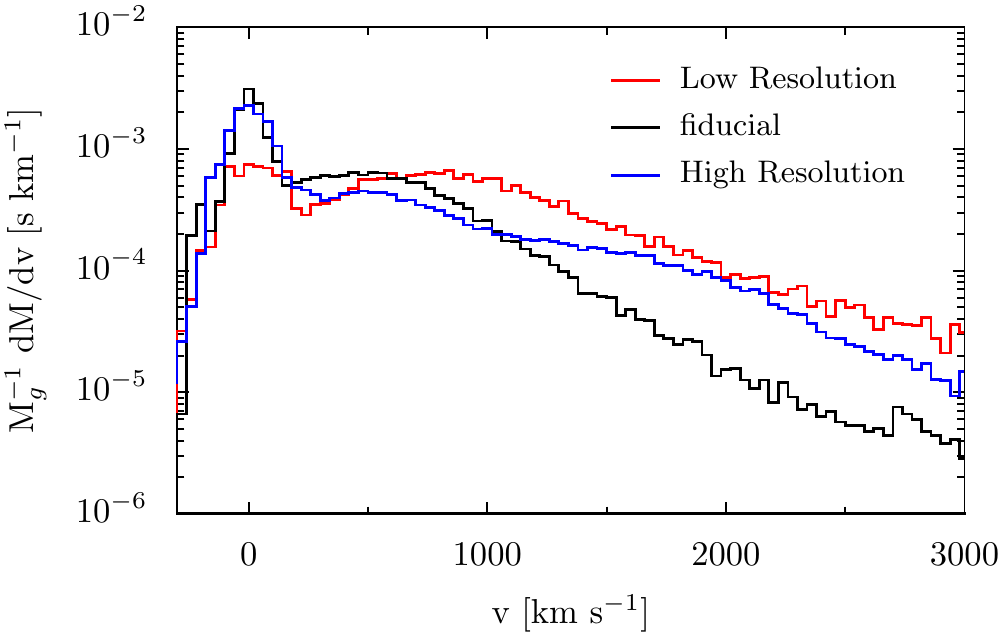}
\caption{ 
The distribution of radial velocities relative to the black hole for the gas
particles in the fiducial simulation (black), the simulation at lower resolution
(red) and the simulation at higher resolution (blue).
}
\label{fig:resolvpdf}
\end{figure}

\begin{table}
 \caption{Resolution Test}
 \label{tab:reso}
 \begin{tabular}{lccc}
  \hline 
  Property & High & Fid & Low \\
  \hline
  $M_{*,t}$ [$10^{9} \msun$] & 11.5 & 10.9 & 10.2 \\
  $M_{BH,f}$ [$10^7 \msun$] & 4.77 & 4.38 & 5.28 \\
  $f_{\rm disk}$ & 0.012 & 0.006 & 0.004 \\
  $f_{\rm out}$ & 0.166 & 0.196 & 0.307 \\
  $M_{\rm out}$ [$10^{9} \msun$] & 2.65 & 3.13 & 4.89 \\ 
 \end{tabular}
 
 \medskip
 Simulation properties are the same as those of the fiducial simulation in Table~1.
 The three simulations differ only in the mass resolution, with a total of $1.6\times 10^5$ (low), $4.8\times 10^5$ (fid), and $1.6 \times 10^6$ (high) particles.
\end{table}

Figure~\ref{fig:resolvpdf} shows the radial velocities of the SPH particles
(relative to the BH) at the end of the simulation for three runs that
differ only in the mass resolution with 1.6, 4.8, and 16.0 $\times 10^5$
particles, respectively.  For the two higher resolution simulations, the distributions
are quite similar for the bulk of the disk ($v \sim 0$) and outflow ($v
\sim 500-1000 \kms$), though the high velocity tails differ somewhat.
Table \ref{tab:reso} compares the integrated
quantities shown in Figure \ref{fig:woblowout}.  The biggest difference is
that the highest resolution simulation retains more mass in a nuclear disk
at late times.  However, the mass in the outflow and the fraction of the
mass in the outflow at late times are reasonably converged.  The total mass
of new stars formed and the final BH mass also do not change appreciably
with resolution.  We conclude that the effects of this model are reasonably
well resolved at the fiducial resolution employed.

\section{Discussion and Conclusions}
\label{sec:discussion}

We have carried out three-dimensional SPH simulations of the interaction between a high-speed outflow produced by an AGN and interstellar gas in the AGN's host galaxy.  We show that this interaction can drive a large-scale galactic outflow that in some cases unbinds a significant fraction of the ISM of the host galaxy (see, e.g., \citealt{king05} for closely related arguments).  The AGN-driven galactic winds found here provide a possible explanation for the high velocity outflows observed in some post-starburst galaxies \citep{tremonti07} and for the massive atomic and molecular outflows with $v \sim 1,000 \kms$ seen in local ultra-luminous infrared galaxies (ULIRGs) \citep{feruglio10,chung11,rupke11,sturm11}.   We return to this below.  Our specific calculations assume that AGN activity is triggered by major mergers of galaxies 
(as seen, for example, by \cite{koss2010}), 
but we suspect that the results presented here are  much more general and apply relatively independent of the physical mechanism(s) that drive gas into the central $\sim 100$ pc of galaxies.   The physically-motivated model utilized here for AGN wind feedback is easy to implement (\S \ref{sec:windfeedback}) and could readily be applied in other contexts, including high resolution simulations of galactic nuclei and cosmological simulations of BH growth and evolution.

 Our results provide a quantitative mapping between the properties of an AGN wind in the vicinity of the BH and the resulting large-scale galaxy-wide outflow (Table \ref{tab:simparm} and Fig.~\ref{fig:woblowout}).  We have parametrized the AGN wind at small radii in terms of its speed $v_w$ and momentum flux $\dot p_w = \tau_w L/c$ -- the corresponding energy flux in the wind is $\dot E_w = 0.5 \, \dot p_w \, v_w = 0.5 \, \tau_w (v_w/c) \, L$.    

We find that an AGN wind with $v_w \sim 10,000 \kms$ -- typical of, e.g., broad-absorption line quasars -- produces a galactic outflow with velocities $\sim 1,000 \kms$.    This reduction in velocity is a consequence of the AGN wind sweeping up and driving out $\sim 5-10$ times as much mass as was in the wind initially (e.g., Fig.~\ref{fig:vpdf}).   Quantitatively, we find that the AGN-driven galactic outflow unbinds $\sim 1-5 \times 10^9 M_\odot$ of gas, $\sim 10-40 \%$ of the initial gas in the two merging galaxies (which are Milky Way-like spirals in our calculations).    This is true for a wide range of outflow models, covering $\tau_w \sim 1-10$, $v_w \sim 7,000-10,000 \, \kms$, and with and without the radiation pressure feedback model from \citet{debuhr10,debuhr11}.   Not surprisingly, AGN winds with larger momentum and energy fluxes (larger $\tau_w$ and $v_w$) are more effective at driving galaxy-scale outflows (Fig.~\ref{fig:woblowout}).   

The mass outflow rate from the vicinity of the galaxy is somewhat noisy and thus difficult to directly measure in the simulations.  The average outflow rate relative to the star formation rate can, however, be readily estimated by comparing the total unbound mass ($M_{\rm out}$ in Table \ref{tab:simparm}) with the mass in new stars formed in the simulation ($M_{*,f}$ and $M_{*,p}$ in Table \ref{tab:simparm}, where $M_{*,f}$ is the total mass of new stars formed and $M_{*,p}$ is the mass of new stars formed after the peak of star formation).   This comparison implies that the average mass outflow rate is typically comparable to the star formation rate; after the peak of star formation in the merger, which is when the BH reaches its final mass, the average outflow rate often exceeds the corresponding star formation rate by factors of a few.  

Physically, we find that the properties of the galaxy-wide outflow are determined by both the momentum and energy fluxes in the AGN wind at small radii.   A larger momentum flux implies that more mass can be swept out of the galaxy while a larger energy flux in the wind (e.g., larger $v_w$ at fixed $\dot p_w$) leads to more shock heating of the ambient ISM.   This shocked gas partially cools but retains enough thermal energy to contribute to driving the galaxy-wide outflow.   For quantitatively accurate results we find that it is necessary to modify the \citet{springel03} equation of state to properly include metal-line cooling and Compton heating/cooling as processes by which gas relaxes back to the effective equation of state that is used to calculate the ISM pressure (see \S \ref{sec:cooling}).

To produce a galaxy at the end of the simulation that is on the $M_{BH}-\sigma$ relation using only feedback by our AGN wind model, we require $\tau_w \gtrsim 5-10$ (see Table~\ref{tab:simparm} and Fig.~\ref{fig:woblowout}; the velocity dispersion of the remnant galaxy is $\simeq 170 \kms$ so that $M_{BH} \simeq 7 \times 10^7 M_\odot$ for galaxies on the mean $M_{BH}-\sigma$ relation).   This required momentum flux is comparable to, although somewhat smaller than, the required momentum flux of $\dot p \sim 20 \, L/c$ that DQM found  using a simple model of radiation pressure feedback in which all particles in the vicinity of the BH feel a force $\propto \tau \, L/c$.   The energy flux corresponding to $\tau_w \sim 5-10$ and $v_w \sim 10,000 \kms$ is $\dot E \sim 0.1 \, L$, similar to the energy injection rate of $\sim 0.05 \, L$ that \citet{dimatteo05} found was required to reproduce the $M_{BH}-\sigma$ relation (although they deposited thermal energy into ambient gas while we supply kinetic energy to a wind).  

One  important distinction that we find is between AGN driving galactic outflows and AGN outflows regulating the growth of the BH.   For example, in our models with $\tau_w = 1$ and $v_w = 10,000 \kms$, the final BH mass is a factor of $\sim 10$ above the mean $M_{BH}-\sigma$ relation in calculations that only include AGN wind feedback.   Also including  the simple radiation pressure feedback model from DQM in the simulation resolves this discrepancy; the same AGN wind model then unbinds $\sim 10^9 M_\odot$ of gas, $\sim 10 \%$ of the initial gas mass in the system.  However, in spite of this outflow, the AGN wind neither regulates the BH growth nor has a significant effect on the average star formation in the system.   This demonstrates that very powerful AGN outflows with $\tau_w \sim 3-10$ are required for the BH to have a substantial effect on either its own accretion history or the star formation in the surrounding galaxy. This conclusion is similar to that of \cite{silk10}, who required a comparable momentum to eject gas from their model galaxies, and to that of \cite{kaviraj11}, who found that simple models for the color evolution of early type galaxies required feedback similar in strength ($\dot{E}$ of a few percent) and duration ($\sim 200$ Myr) to what we find here.

The galactic outflows found in our simulations with $v_w \sim 10,000 \kms$ and $\tau_w \gtrsim 3$ have properties broadly similar to those observed in some massive post-starburst galaxies \citep{tremonti07} and in local ULIRGs \citep{feruglio10,chung11,rupke11,sturm11}.   Moreover, it is striking that \citet{sturm11} inferred $\dot p_w \sim 10  \, L/c$, i.e., $\tau_w \sim 10$, for the molecular outflows in local ULIRGs (these values are uncertain at the factor of $\sim 3$ level; see, e.g., \citealt{sturm11} for more details).  This is comparable to the outflow momentum flux we find is required to {both} suppress late time star formation during galaxy mergers and produce remnants approximately on the observed $M_{BH}-\sigma$ relation. This correspondence is encouraging, though it does not directly address the question of what mechanism powers such outflows in the first place.   The relative role of star formation and AGN in  powering the $\sim 1000 \kms$ molecular outflows is also  unclear, with \citet{sturm11} presenting evidence from OH observations in favor of AGN, while \citet{chung11} argue that star-formation is the dominant power source based on their CO observations.  In addition, in some samples of post-starburst galaxies, the observed outflows appear consistent with those driven by star formation \citep{coil11}.

From the theoretical perspective, the biggest uncertainty in applying our results is the uncertainty in the absolute momentum/energy flux in AGN outflows at small radii.   This uncertainty remains even for the most well-characterized outflows, those of broad-absorption line (BAL) quasars.  In theoretical models in which BAL outflows are produced by radiation pressure on lines in the vicinity of the broad-line region, the predicted momentum flux is typically  $\lesssim  L/c$, i.e., $\tau_w \lesssim 1$ \citep{murray1995}.  This follows  from the fact that (observationally) the lines  do not cover the entire continuum and thus cannot absorb the entire continuum momentum flux.   This constraint is less clear for low-ionization BALs (in which the lines cover more of the continuum), but probably remains reasonably accurate.   One way out of this conclusion is if magnetic fields contribute to accelerating BAL outflows (as in, e.g., \citealt{proga03}) since this can increase both the mass flux and terminal velocity of the outflow.    There are observational estimates that the momentum flux in some low-ionization BAL outflows may indeed be $\gtrsim L/c$  \citep{moe09, bautista10,dunn10}, though these are difficult observations to interpret and they have complicated selection effects.   

In addition to line-driven winds, it is possible that even more powerful outflows are driven during evolutionary phases distinct from optically bright quasars.   The two most promising from our point of view are:  (1) outflows driven by radiation pressure on dust during phases when the ISM around the AGN is optically thick to far-infrared radiation \citep{murray05,debuhr10}.   In this case the momentum flux in a small-scale AGN wind can in principle reach $\sim \tau L/c$, where $\tau$ is the infrared optical depth of the galactic nucleus.  Such an outflow is not directly seen in DQM's simple implementation of radiation pressure feedback.  However, this could easily be a limitation of the lack of a proper treatment of the infrared radiative transfer.  Indeed, we expect that radiation pressure will inevitably drive a high-speed outflow off of the nuclear disk (`torus') at $\sim 1-10$ pc.    (2)   An additional source of powerful high-speed outflows can be produced if the fueling rate of the central AGN significantly exceeds Eddington (e.g., \citealt{king03}).  In this case, the accretion disk becomes radiatively inefficient and much of the nominally inflowing mass is likely unbound (e.g., \citealt{blandford99}).   The disparity in spatial scales between the horizon and where the AGN accretion rate is set ($\sim 1-10$ pc; e.g., \citealt{hopkins10c}) implies that it is very likely that there is a phase of super-Eddington fueling.  The critical question is whether this phase lasts long enough to dominate the {integrated} properties of a black hole's outflow.

All of the calculations presented in this paper utilize the effective equation of state model of \citet{springel03}  (modified as in \S \ref{sec:cooling}) to pressurize the ISM and prevent runaway gravitational fragmentation.   This is a very simplistic treatment of ISM physics and in the future it is important to study the interaction between a central AGN and the surrounding ISM using more realistic  models that approximate the turbulent, multi-phase structure of the ISM (e.g., \citealt{dobbs11,hopkins11}).

\section*{Acknowledgments}

We thank Norm Murray for useful conversations.  Support for EQ and CPM was
provided in part by the Miller Institute for Basic Research in Science,
University of California Berkeley.  CPM is also supported by NASA through
grant 10-ATP10-0187 and grant HST-AR-12140.01-A from the Space Telescope
Science Institute, which is operated by the Association of Universities for
Research in Astronomy, Inc., under contract NAS 5-26555.  The computations
for this paper were performed on the {\em Henyey} cluster at UC Berkeley
supported by NSF AST Grant 0905801, the National Energy Research Scientific
Computing Center supported by the Office of Science of the U.S. Department
of Energy under Contract No. DE-AC02-05CH11231, and the Texas Advanced
Computing Center (TACC) at The University of Texas at Austin.

\bibliographystyle{mn2e}
\bibliography{thesis}

\begin{thebibliography}{}

\bibitem[\protect\citeauthoryear{{Bautista}, {Dunn}, {Arav}, {Korista}, {Moe}
  \& {Benn}}{{Bautista} et~al.}{2010}]{bautista10}
{Bautista} M.~A.,  {Dunn} J.~P.,  {Arav} N.,  {Korista} K.~T.,  {Moe} M.,
  {Benn} C.,  2010, \apj, 713, 25

\bibitem[\protect\citeauthoryear{{Blandford} \& {Begelman}}{{Blandford} \&
  {Begelman}}{1999}]{blandford99}
{Blandford} R.~D.,  {Begelman} M.~C.,  1999, \mnras, 303, L1

\bibitem[\protect\citeauthoryear{{Chung}, {Yun}, {Naraynan}, {Heyer} \&
  {Erickson}}{{Chung} et~al.}{2011}]{chung11}
{Chung} A.,  {Yun} M.~S.,  {Naraynan} G.,  {Heyer} M.,    {Erickson} N.~R.,
  2011, \apjl, 732, L15+

\bibitem[\protect\citeauthoryear{{Coil}, {Weiner}, {Holz}, {Cooper}, {Yan} \&
  {Aird}}{{Coil} et~al.}{2011}]{coil11}
{Coil} A.~L.,  {Weiner} B.~J.,  {Holz} D.~E.,  {Cooper} M.~C.,  {Yan} R.,
  {Aird} J.,  2011, \apj, 743, 46

\bibitem[\protect\citeauthoryear{{Croton}, {Springel}, {White}, {De Lucia},
  {Frenk}, {Gao}, {Jenkins}, {Kauffmann}, {Navarro} \& {Yoshida}}{{Croton}
  et~al.}{2006}]{croton06}
{Croton} D.~J.,  {Springel} V.,  {White} S.~D.~M.,  {De Lucia} G.,  {Frenk}
  C.~S.,  {Gao} L.,  {Jenkins} A.,  {Kauffmann} G.,  {Navarro} J.~F.,
  {Yoshida} N.,  2006, \mnras, 365, 11

\bibitem[\protect\citeauthoryear{{Dai}, {Shankar} \& {Sivakoff}}{{Dai}
  et~al.}{2008}]{dai08}
{Dai} X.,  {Shankar} F.,    {Sivakoff} G.~R.,  2008, \apj, 672, 108

\bibitem[\protect\citeauthoryear{{DeBuhr}, {Quataert} \& {Ma}}{{DeBuhr}
  et~al.}{2011}]{debuhr11}
{DeBuhr} J.,  {Quataert} E.,    {Ma} C.,  2011, \mnras, 412, 1341

\bibitem[\protect\citeauthoryear{{DeBuhr}, {Quataert}, {Ma} \&
  {Hopkins}}{{DeBuhr} et~al.}{2010}]{debuhr10}
{DeBuhr} J.,  {Quataert} E.,  {Ma} C.,    {Hopkins} P.,  2010, \mnras, 406, L55

\bibitem[\protect\citeauthoryear{{Di Matteo}, {Springel} \& {Hernquist}}{{Di
  Matteo} et~al.}{2005}]{dimatteo05}
{Di Matteo} T.,  {Springel} V.,    {Hernquist} L.,  2005, \nat, 433, 604

\bibitem[\protect\citeauthoryear{{Dobbs}, {Burkert} \& {Pringle}}{{Dobbs}
  et~al.}{2011}]{dobbs11}
{Dobbs} C.~L.,  {Burkert} A.,    {Pringle} J.~E.,  2011, \mnras, 417, 1318

\bibitem[\protect\citeauthoryear{{Dunn}, {Bautista}, {Arav}, {Moe}, {Korista},
  {Costantini}, {Benn}, {Ellison} \& {Edmonds}}{{Dunn} et~al.}{2010}]{dunn10}
{Dunn} J.~P.,  {Bautista} M.,  {Arav} N.,  {Moe} M.,  {Korista} K.,
  {Costantini} E.,  {Benn} C.,  {Ellison} S.,    {Edmonds} D.,  2010, \apj,
  709, 611

\bibitem[\protect\citeauthoryear{{Durier} \& {Dalla Vecchia}}{{Durier} \&
  {Dalla Vecchia}}{2011}]{durier11arxiv}
{Durier} F.,  {Dalla Vecchia} C.,  2011, ArXiv e-prints

\bibitem[\protect\citeauthoryear{{Faucher-Giguere}, {Quataert} \&
  {Murray}}{{Faucher-Giguere} et~al.}{2011}]{claude11}
{Faucher-Giguere} C.~.,  {Quataert} E.,    {Murray} N.,  2011, ArXiv e-prints

\bibitem[\protect\citeauthoryear{{Feruglio}, {Maiolino}, {Piconcelli}, {Menci},
  {Aussel}, {Lamastra} \& {Fiore}}{{Feruglio} et~al.}{2010}]{feruglio10}
{Feruglio} C.,  {Maiolino} R.,  {Piconcelli} E.,  {Menci} N.,  {Aussel} H.,
  {Lamastra} A.,    {Fiore} F.,  2010, \aap, 518, L155+

\bibitem[\protect\citeauthoryear{{Hernquist}}{{Hernquist}}{1990}]{hernquist90}
{Hernquist} L.,  1990, \apj, 356, 359

\bibitem[\protect\citeauthoryear{{Hopkins} \& {Quataert}}{{Hopkins} \&
  {Quataert}}{2010}]{hopkins10c}
{Hopkins} P.~F.,  {Quataert} E.,  2010, \mnras, 407, 1529

\bibitem[\protect\citeauthoryear{{Hopkins}, {Quataert} \& {Murray}}{{Hopkins}
  et~al.}{2011}]{hopkins11}
{Hopkins} P.~F.,  {Quataert} E.,    {Murray} N.,  2011, \mnras, 417, 950

\bibitem[\protect\citeauthoryear{{Kaviraj}, {Schawinski}, {Silk} \&
  {Shabala}}{{Kaviraj} et~al.}{2011}]{kaviraj11}
{Kaviraj} S.,  {Schawinski} K.,  {Silk} J.,    {Shabala} S.~S.,  2011, \mnras,
  415, 3798

\bibitem[\protect\citeauthoryear{{King}}{{King}}{2003}]{king03}
{King} A.,  2003, \apjl, 596, L27

\bibitem[\protect\citeauthoryear{{King}}{{King}}{2005}]{king05}
{King} A.,  2005, \apjl, 635, L121

\bibitem[\protect\citeauthoryear{{King} \& {Pounds}}{{King} \&
  {Pounds}}{2003}]{kingpounds03}
{King} A.~R.,  {Pounds} K.~A.,  2003, \mnras, 345, 657

\bibitem[\protect\citeauthoryear{{King}, {Zubovas} \& {Power}}{{King}
  et~al.}{2011}]{king11}
{King} A.~R.,  {Zubovas} K.,    {Power} C.,  2011, \mnras, 415, L6

\bibitem[\protect\citeauthoryear{{Koss}, {Mushotzky}, {Veilleux} \&
  {Winter}}{{Koss} et~al.}{2010}]{koss2010}
{Koss} M.,  {Mushotzky} R.,  {Veilleux} S.,    {Winter} L.,  2010, \apjl, 716,
  L125

\bibitem[\protect\citeauthoryear{{Krolik}}{{Krolik}}{1999}]{krolik99}
{Krolik} J.~H.,  1999, {Active galactic nuclei : from the central black hole to
  the galactic environment}

\bibitem[\protect\citeauthoryear{{Moe}, {Arav}, {Bautista} \& {Korista}}{{Moe}
  et~al.}{2009}]{moe09}
{Moe} M.,  {Arav} N.,  {Bautista} M.~A.,    {Korista} K.~T.,  2009, \apj, 706,
  525

\bibitem[\protect\citeauthoryear{{Murray}, {Chiang}, {Grossman} \&
  {Voit}}{{Murray} et~al.}{1995}]{murray1995}
{Murray} N.,  {Chiang} J.,  {Grossman} S.~A.,    {Voit} G.~M.,  1995, \apj,
  451, 498

\bibitem[\protect\citeauthoryear{{Murray}, {Quataert} \& {Thompson}}{{Murray}
  et~al.}{2005}]{murray05}
{Murray} N.,  {Quataert} E.,    {Thompson} T.~A.,  2005, \apj, 618, 569

\bibitem[\protect\citeauthoryear{{Novak}, {Ostriker} \& {Ciotti}}{{Novak}
  et~al.}{2011}]{novak10}
{Novak} G.~S.,  {Ostriker} J.~P.,    {Ciotti} L.,  2011, \apj, 737, 26

\bibitem[\protect\citeauthoryear{{Ostriker}, {Choi}, {Ciotti}, {Novak} \&
  {Proga}}{{Ostriker} et~al.}{2010}]{ostriker10}
{Ostriker} J.~P.,  {Choi} E.,  {Ciotti} L.,  {Novak} G.~S.,    {Proga} D.,
  2010, \apj, 722, 642

\bibitem[\protect\citeauthoryear{{Price}}{{Price}}{2007}]{price07}
{Price} D.~J.,  2007, Publications of the Astronomical Society of Australia,
  24, 159

\bibitem[\protect\citeauthoryear{{Proga}}{{Proga}}{2003}]{proga03}
{Proga} D.,  2003, \apj, 585, 406

\bibitem[\protect\citeauthoryear{{Rupke} \& {Veilleux}}{{Rupke} \&
  {Veilleux}}{2011}]{rupke11}
{Rupke} D.~S.~N.,  {Veilleux} S.,  2011, \apjl, 729, L27+

\bibitem[\protect\citeauthoryear{{Sazonov}, {Ostriker} \& {Sunyaev}}{{Sazonov}
  et~al.}{2004}]{sazonov04}
{Sazonov} S.~Y.,  {Ostriker} J.~P.,    {Sunyaev} R.~A.,  2004, \mnras, 347, 144

\bibitem[\protect\citeauthoryear{{Sharma}, {Parrish} \& {Quataert}}{{Sharma}
  et~al.}{2010}]{sharma10}
{Sharma} P.,  {Parrish} I.~J.,    {Quataert} E.,  2010, \apj, 720, 652

\bibitem[\protect\citeauthoryear{{Silk} \& {Nusser}}{{Silk} \&
  {Nusser}}{2010}]{silk10}
{Silk} J.,  {Nusser} A.,  2010, \apj, 725, 556

\bibitem[\protect\citeauthoryear{{Silk} \& {Rees}}{{Silk} \&
  {Rees}}{1998}]{silk98}
{Silk} J.,  {Rees} M.~J.,  1998, \aap, 331, L1

\bibitem[\protect\citeauthoryear{{Springel}}{{Springel}}{2005}]{springel05}
{Springel} V.,  2005, \mnras, 364, 1105

\bibitem[\protect\citeauthoryear{{Springel} \& {Hernquist}}{{Springel} \&
  {Hernquist}}{2003}]{springel03}
{Springel} V.,  {Hernquist} L.,  2003, \mnras, 339, 289

\bibitem[\protect\citeauthoryear{{Sturm}, {Gonz{\'a}lez-Alfonso}, {Veilleux},
  {Fischer} \& {Graci{\'a}-Carpio}}{{Sturm} et~al.}{2011}]{sturm11}
{Sturm} E.,  {Gonz{\'a}lez-Alfonso} E.,  {Veilleux} S.,  {Fischer} J.,
  {Graci{\'a}-Carpio} J. e.~a.,  2011, \apjl, 733, L16+

\bibitem[\protect\citeauthoryear{{Sutherland} \& {Dopita}}{{Sutherland} \&
  {Dopita}}{1993}]{sutherland93}
{Sutherland} R.~S.,  {Dopita} M.~A.,  1993, \apjs, 88, 253

\bibitem[\protect\citeauthoryear{{Tombesi}, {Cappi}, {Reeves}, {Palumbo},
  {Yaqoob}, {Braito} \& {Dadina}}{{Tombesi} et~al.}{2010}]{tombesi10}
{Tombesi} F.,  {Cappi} M.,  {Reeves} J.~N.,  {Palumbo} G.~G.~C.,  {Yaqoob} T.,
  {Braito} V.,    {Dadina} M.,  2010, \aap, 521, A57+

\bibitem[\protect\citeauthoryear{{Tremonti}, {Moustakas} \&
  {Diamond-Stanic}}{{Tremonti} et~al.}{2007}]{tremonti07}
{Tremonti} C.~A.,  {Moustakas} J.,    {Diamond-Stanic} A.~M.,  2007, \apjl,
  663, L77

\end{thebibliography}

\end{document}